\pdfoutput=1

\documentclass[11pt]{article}

\usepackage[final]{acl}

\usepackage{times}
\usepackage{latexsym}
\usepackage[T1]{fontenc}
\usepackage[utf8]{inputenc}
\usepackage{microtype}
\usepackage{inconsolata}
\usepackage{graphicx}

\title{Identifying High Consideration E-Commerce Search Queries}

\author{Zhiyu Chen ~~~~ Jason Choi ~~~~ Besnik Fetahu ~~~~ Shervin Malmasi \\
 Amazon.com, Inc. ~~~ Seattle, WA, USA \\
\texttt{\{zhiyuche,chojson,besnikf,malmasi\}@amazon.com}}

\usepackage{hyperref}
\usepackage{amsthm}
\usepackage{color,soul}
\usepackage{enumitem}
\usepackage{cleveref}
\usepackage{tabularx}
\usepackage{graphicx}
\usepackage{subcaption}
\usepackage{svg}
\usepackage{colortbl}
\usepackage{booktabs}
\usepackage{amsfonts}

\definecolor{darkred}{RGB}{139,0,0}
\definecolor{verylightred}{RGB}{255,243,240}
\begin{document}
\maketitle
\begin{abstract}
In e-commerce, high consideration search missions typically require careful and elaborate decision making, and involve a substantial research investment from customers. 
We consider the task of automatically identifying such High Consideration (HC) queries.
Detecting such missions or searches enables e-commerce sites to better serve user needs through targeted experiences such as curated QA widgets that help users reach purchase decisions. 
We explore the task by proposing an Engagement-based Query Ranking (EQR) approach, focusing on query ranking to indicate potential engagement levels with query-related shopping knowledge content during product search. Unlike previous studies on predicting trends, EQR prioritizes query-level features related to customer behavior, financial indicators, and catalog information rather than popularity signals.
We introduce an accurate and scalable method for EQR and present experimental results demonstrating its effectiveness. Offline experiments show strong ranking performance. Human evaluation shows a precision of 96\% for HC queries identified by our model. The model was commercially deployed, and shown to outperform human-selected queries in terms of downstream customer impact, as measured through engagement.
\end{abstract}

\begin{figure}[ht]
    \centering

  \begin{subfigure}[b]{1\columnwidth}
        \centering
        \includegraphics[width=0.98\columnwidth]{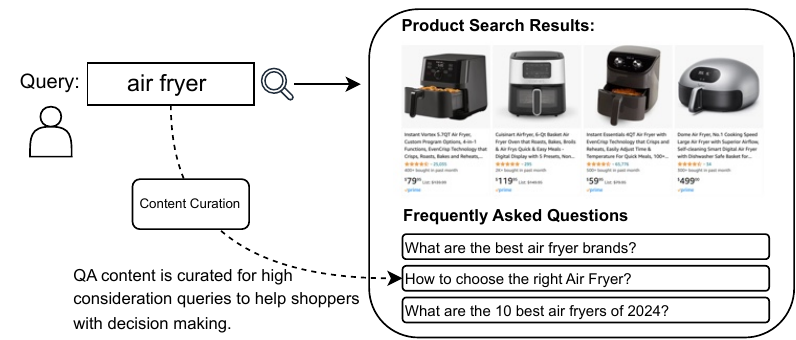}
        \caption{An example of a QA component in search results for the high-engagement query ``air fryer''.}
        \label{fig:motivation}
    \end{subfigure}
    
    \vskip\baselineskip

     \begin{subfigure}[b]{1\columnwidth}
        \centering
        \includegraphics[width=1\columnwidth]{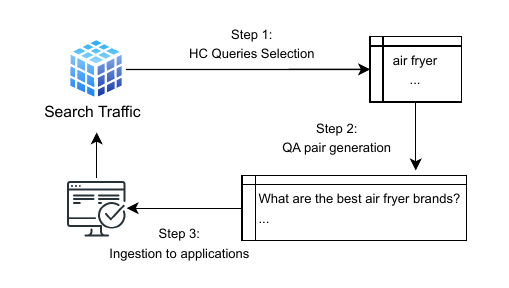}
        \caption{Content curation process for a question answering widget.}
        \label{fig:curation}
    \end{subfigure}

    \caption{An end-to-end demonstration of how the question-answer pairs of a widget is curated.}
    \label{fig:overall}
\end{figure}

\section{Introduction}
\label{sec:intro}

The integration of information content in online shopping is increasingly gaining importance \cite{vedula2024question,kuzi2024bridging}, but such content may be more useful for certain search missions than others.
The effective identification of specific subsets of keywords~\cite{zhao2019dynamic, yuan2022community, ryali2023trendspotter} is essential not only for driving organic traffic and revenue in e-commerce, but also for enhancing the overall customer experience.
Related to the tasks of keyword selection and targeting \cite{shu2020joint, zheng2020predicting}, creating or serving customized content for specific queries helps provide relevant content for the right population.
For example, when customers search for ``prime day deals'' on Amazon, a dedicated widget will appear above product search results.
Clicking on this widget will direct customers to a specialized page with customized information, thus enhancing their shopping experience for specific deals. 
Identifying such queries is usually the initial step prior to content creation and targeting.
Rather than directing customers to a dedicated web page, such targeted content could also be co-displayed with product search results when customers are looking for products. For example, a Question-Answer (QA) widget with relevant shopping knowledge could be rendered to match the customer's query, as shown in \Cref{fig:motivation}.

\pagebreak

Such tailored content is generally most useful for purchases requiring exploration, comparison, and deep decision making \cite{sondhi2018taxonomy}.
We refer to such searches as \textbf{\textit{High-Consideration (HC) Queries}} since consumers require additional information to consider their decision, or refine their search \cite{branco2012optimal}.
However, curating customized content is an expensive manual process, and not feasible for all shopping queries as the query space is in the hundreds of millions. 
In Figure~\ref{fig:curation}, we illustrate the end-to-end process of creating curated content. Initially, HC queries are selected, typically through a manual process guided by heuristics. This often involves human annotators reviewing the top queries ranked by search frequency and identifying potential HC queries based on subjective criteria. Next, customized content is manually curated for each selected query. To serve a QA widget, corresponding question-answer pairs are then created. Finally, these customized QA pairs are ingested into a database and retrieved for specific queries~\cite{chen-etal-2023-generate} as in \Cref{fig:motivation}.

Manual keyword selection by humans is the most straightforward approach. 
However, even with frequency-ranked query lists this is an expensive and low-yield process as most queries are judged to be of low consideration (e.g., consumables and minor purchases). 
Therefore, a lot of human efforts can be wasted using the conventional method. 
To address this, our work focuses on the task of automatically identifying the small subset of such HC queries in this large space. 
Identifying the most valuable queries (i.e., step 1 in Figure~\ref{fig:curation}) is crucial for maximizing the Return on Investment (ROI) of human efforts dedicated to content curation (i.e., step 2 in Figure~\ref{fig:curation}).
As discussed later, the cost associated with content creation (and other factors), is an important consideration in framing this as a ranking task rather than a classification task.

We hypothesize that HC queries can be identified by a combination of behavioral cues, financial signals, and catalog features. 
To identify HC queries, we propose the novel task of \textbf{\textit{Engagement-based Query Ranking}} (EQR) to train a model leveraging these signals. 
To learn this ranking function, our approach relies on engagement with informational shopping content in search results (e.g. a QA element) as a proxy target for HC query labels (which are expensive to define).
This engagement can be measured as the Click-Through Rate (CTR) of the content displayed for a set of seed queries. 
As we will show, these targets can then be used to create a generalizable model to predict HC queries across all search traffic. 
This approach also allows for continuous learning by using engagement from new content created for queries selected by our model.

Our key contributions are summarized below:
\begin{itemize}[leftmargin=*]
  \setlength{\itemsep}{1pt}
  \setlength{\parskip}{0pt}
  \setlength{\parsep}{0pt}
    \item We introduce a novel task called Engagement-based Query Ranking (EQR), aimed at ranking HC queries based on their engagement metrics.
    \item We propose a simple and effective method for EQR, which could effective identify novel queries that may result in prospective future engagement.
    \item Our offline experiments show our proposed method outperforms all baselines for EQR across all metrics. And a human evaluation measured the model's precision at 96\% in terms of HC queries selection. 
    \item Finally, commercial deployment of the model showed that the downstream customer impact from its selected HC queries is higher than those selected by human annotators.
\end{itemize}

\section{Related Work}

\paragraph{\textbf{Query Understanding in E-commerce}} 
Query understanding is important to optimize search results for e-commerce platforms. To improve the relevance of search results while preserving the recall, embedding-based methods~\cite{lin2018commerce} have been proposed to map a query into a target product category. The first empirical study on e-commerce queries was conducted by \citet{sondhi2018taxonomy} and they categorized e-commerce queries into five categories based on different search behaviors. 
\citet{chen-etal-2023-generate} introduced an intent classifier to determine whether to display an FAQ entry for a given query. Our work can be considered as an extension to \citet{chen-etal-2023-generate} and focuses on identifying new queries where customers could benefit from the associated content from a QA component. Once those queries are identified, we could expand the QA database accordingly to increase its coverage.

\paragraph{\textbf{Query Performance Prediction}}
In information retrieval, the task of query performance prediction (QPP)~\cite{carmel2012query} aims to predict the effectiveness of a query given a retrieval system without using human-labeled relevance judgments. QPP methods can generally be categorized into pre-retrieval and post-retrieval methods~\cite{hauff2008survey}. 
Pre-retrieval methods are designed to estimate query performance before the retrieval stage is reached, and they utilize various features such as query term characteristics and collection statistics to make their predictions~\cite{mothe2005linguistic}. 
On the other hand, post-retrieval techniques~\cite{cronen2002predicting, roitman2019query} focus on deriving predictions from the ranked list of results obtained through the retrieval process. Unlike pre-retrieval methods, post-retrieval predictors have access to the actual search outcomes, which can provide valuable additional information for analysis. For example, Query Clarity~\cite{cronen2002predicting} evaluates the quality of search results by measuring the KL divergence between language models derived from the search results and those from the corpus. 
For the first time, \citet{kumar2018did} performed query performance prediction in e-commerce domain. 

Our method for identifying high consideration queries shares commonalities with QPP methods, which provide insights for a query without relying on human judgments. Instead of focusing on retrieval quality for a given query, we care about the downstream business impact (e.g., click-through rate) of curated content for a selected query. We also model the task as a regression task to predict a target measure of queries~\cite{zamani2018neural, hashemi2019performance,arabzadeh2021bert,khodabakhsh2023learning}.

\paragraph{\textbf{Trending Queries Detection}} 
Our work is also related to \textit{detection of trending queries}. 
\citet{giummole2013trending} analyzed on real data and showed that a topic trending on Twitter may subsequently emerge as a popular search query on Google. 
\citet{10.1145/2567948.2577315} proposed to predict trending queries with a classifier trained on features derived from the historical frequencies of queries. 
More recently, trending prediction has also been explored in e-commerce scenario. \citet{yuan2022community} introduced a method to mine fashion trends represented by product attributes on e-commerce platforms. TrendSpotter was proposed by \citet{ryali2023trendspotter} to forecast trending products.
Different from prior work, our goal is to identify high-consideration queries that could encourage user engagement with related content, rather than focusing on the queries themselves.
A trending query may not necessarily fall into the high-consideration category due to specific business considerations. Additionally, our method does not rely on surface features of queries and we aim to discover new HC queries.

\begin{table}[t!]
\resizebox{0.99\columnwidth}{!}{
\begin{tabular}{cp{10cm}}
\toprule
\textbf{Feature} & \textbf{Description} \\
\midrule
B1 & The average number of add-to-cart actions attributed to the query. \\
B2 & Average daily search count in last 30 days. \\
B3 & Average number of add-to-cart actions after a search (100-minute window). \\
B4 & Average number of clicks after a search (100-minute window). \\
B5 & Average ranking of the first result clicked in the search. \\
B6 & The average 30 days add-to-cart rate of the search query. \\
B7 & Time elapsed from the search to the first add-to-cart action. \\ 
B8 & The average count of viewed products in search results.\\ \midrule
F1 & Daily average product sales of the search query. \\
F2 & The average product sales within a 100-minute window after the first search of the query in the same session. \\
F3 & The average product sales value from all purchases made following a search occurs on the same day. \\
F4 & The average product sales value attributed to the query. \\
F5 & Average product sales value from purchases of products that were sponsored on the search results of the query. \\
\midrule
C1 & The average number of results found for a query. \\
C2 & The average number of results displayed for a query. \\
C3 & The average number of products shown that are sponsored. \\
\bottomrule
\end{tabular}
}\caption{Description of our proposed features for EQR.}
\label{t:feature}
\end{table}

\section{Method}

\subsection{Classification and Ranking Approaches}
\label{sec:rank-or-classify}

Since we aim to predict a subset of queries according to our criteria, this task could be framed as either classification or ranking.
However, the classification approach is overly simplistic, and we model the task of HC queries identification as a ranking task instead of classification task for the following considerations.

\paragraph{Shortcomings of Binary Classification}

Ideally, we want to create tailored content for every targeted query. 
However, not all queries will have content engagement. User needs for informational content are subjective and depend on factors such as knowledge level. By taking a ranking approach we can identify HC queries with higher engagement and prioritize them, which is an important aspect given the constrained human resources required for content creation and validation. 
A simple classification approach would not allow us to maximize the impact of our efforts in production by addressing the most promising queries first. 
Additionally, using a classification approach would require additional steps for threshold selection, as well as ensuring that probability outputs from the model are well calibrated, both of which can be avoided with a ranking approach.

\paragraph{Ranking Approach}
Since our primary objective is scoring queries independently, rather than relative ranking, we choose to model our task as a pointwise ranking problem, similar to the work on query performance prediction~\cite{zamani2018neural, hashemi2019performance,arabzadeh2021bert,khodabakhsh2023learning, meng2024query}. 
This allows us to train a model using data with absolute engagement scores, rather than collecting pairwise comparisons. 
A pointwise approach is also preferred from a model complexity perspective. It allows us to use standard regression approaches, rather than more sophisticated pairwise or listwise ranking models. This is preferred as the lower computational cost leads to faster training and inference.
A pointwise approach is also more interpretable, and makes it easy to perform inference on new queries as they appear.
Since it is trained on engagement data, it is also the most suitable for incremental or online learning, allowing the pointwise model to easily adapt to changing user behavior.
Accordingly, a pointwise approach is the most scalable and practical solution.

\subsection{Pointwise Ranking Model}
\label{sec:model}

Our goal is to create a supervised ranking model for selecting HC queries. 
We first define our proxy target, the query ($q$) level engagement score ($e$), as following:
\begin{equation}\label{eq:y}
    e(q) = \frac{freq_c(q)}{freq(q)}
\end{equation}
where $freq_c(q)$ is the user click count for an informational component (e.g. QA widget like in \Cref{fig:motivation}) displayed in the search results, and $freq(q)$ is the total frequency for the query. Note that the definition of the engagement score can vary among different businesses. 
In this work, we consider the overall query-level engagement instead of content-level engagement like the CTR of an individual QA pair generated for a query. The scope of this paper is on the selection of HC queries and we leave the study of how to guide the content generation for optimizing CTR as future work. 

For a query $q \in Q$, we have its user interaction features $x \in \mathbb{R}^d$ where $d$ is the total number of features. 
We aim to learn a function $f(\cdot)$ that could predict the engagement score of a query given its features.
Once we obtain the predicted engagement scores for a list of \(n\) queries \(Q = \{q_1, \ldots, q_i, \ldots, q_n\}\), then we re-rank them in descending order of their engagement scores.

We do not rely on query embeddings for several reasons. 
First, the query space is large, and differences between queries can be nuanced (\textit{airpods} vs. \textit{airpods case}), possibly leading to poor generalization. As we will demonstrate in Section~\ref{sec:rs}, embedding-based methods perform inferior compared to our proposed method, which does not rely on query surface text features. 
Second, text encoders are slower in both training and inference. 
Finally, by using behavioral features our model is language agnostic without using a multilingual encoder.

\paragraph{Training Data Acquisition}
In order to bootstrap the model training, a key requirement is to have a small but representative set of seed queries with engagement scores to train a model that can generalize to previously untargeted queries.\footnote{This includes queries that are unseen in the model training set but present in search traffic, or queries that were never selected as HC queries for a target widget (i.e., the QA component in our case).}
In general this seed query set should be manually chosen by experts, and be stratified over product categories for coverage.

Next, we summarize the features we used (\Cref{sec:feature}) and then describe how we train the ranking model (\Cref{sec:train}).

\subsection{Query Features}
\label{sec:feature}

We have three feature groups, and all features are listed in \Cref{t:feature}.

\textbf{\textit{Behavioral Features}} (B1-B8) characterize user interactions with a search system for a query. After a query is submitted, we observe subsequent interactions such as clicking on a search result, or adding a product to the cart.
We hypothesize that how users interact with the results (e.g. number of item clicks, going deeper into the results, etc.) can help identify HC queries. 

\textbf{\textit{Financial Features}} (F1-F5) relate to the purchases associated with a query. 
We hypothesize that financial signals (e.g. order volumes, prices, temporal patterns) can help distinguish HC queries. 

\textbf{\textit{Catalog Features}} (C1-C3) focus on features of the products served in the search results, as derived from the product catalog. Such features serve as feedback from the product search system and are inspired by post-retrieval methods for predicting query performance~\cite{cronen2002predicting, roitman2019query, butman2013query}.

\subsection{Training}
\label{sec:train}

We approach the EQR task as pointwise regression and train Gradient Boosted Decision Trees (GBDT) to predict query engagement scores. During training we minimize the Mean Squared Error (MSE):
\begin{equation}
    \mathcal{L} = \frac{1}{2}\sum_{i} (f(x_i) - e(q_i))^2
\end{equation}
where $x_i$ is the feature vector of query $q_i$.
In this paper, we adopt the XGBoost~\cite{chen2016xgboost} implementation to train our models.

\section{Experimental Setup}\label{sec:setup}

\paragraph{Dataset}
As illustrated in \Cref{fig:motivation}, we collected the data associated with a QA component from Amazon spanning a one-year period. 
During this time, 11,273 queries were manually selected to trigger the QA component and display curated content. We obtained the corresponding query-level user engagement data as our proxy targets. The data was divided into training, validation, and test sets, with respective proportions of 70\%, 15\%, and 15\%.

\subsection{Evaluation Metrics}

To evaluate the performance of different methods, we use the following evaluation metrics:
\begin{itemize}[leftmargin=*]
  \setlength{\itemsep}{1pt}
  \setlength{\parskip}{0pt}
  \setlength{\parsep}{0pt}
    \item \textbf{HIT$@k$}:  Considering $k$ queries with the highest ground-truth engagement scores as positives, this is the ratio of positive queries in top-$k$ predicted results by the model.

    \item \textbf{Kendall's Tau}: An ordinal rank correlation coefficient for two lists (our predictions and ground truth). Values lie in [-1, 1], and larger values indicate greater similarity \cite{kendall1938new}. This metric is also commonly used in query performance prediction~\cite{hauff2009query}.

    \item \textbf{MSE}: Mean Square Error between the ground-truth engagement and predictions. It is a more challenging metric as it requires capturing the precise engagement levels, which may fluctuate due to seasonal variations.
\end{itemize}

The HIT@$k$ metric assesses the ranking performance for top queries, while Kendall's Tau coefficient evaluates the performance across the entire test set.
While the former two metrics focus on query ranking performance, MSE measures exact engagement prediction.

\subsection{Baselines}

We compare our GBDT ranker with the following methods.

\begin{itemize}[leftmargin=*]
  \setlength{\itemsep}{1pt}
  \setlength{\parskip}{0pt}
  \setlength{\parsep}{0pt}
    \item \textit{Frequency}: Queries are ranked by frequency. This baseline measures whether popularity is a predictor for detecting HC queries.  
    \item \textit{Regression methods}: We use the features outlined in Section~\ref{sec:feature} for both Random Forest and linear models, which include Lasso, Ridge, Elastic Net, and linear regression.
    \item \textit{RoBERTa}: A 300M parameter encoder pre-trained on internal shopping data. We fine-tune this with our training data.
\end{itemize}
We employed a grid search on the validation set to select hyperparameters for all methods.

Inspired by the work of using LLMs for query performance prediction~\cite{meng2024query}, we also adopt LLMs to predict the engagement scores (i.e., \cref{eq:y}) of queries in our testing set. 
\begin{itemize}[leftmargin=*]
  \setlength{\itemsep}{1pt}
  \setlength{\parskip}{0pt}
  \setlength{\parsep}{0pt}
    \item \textit{GPT-3.5 and GPT-4o (zero-shot)}: We developed a prompt to follow our task definition without any examples.
    \item \textit{GPT-3.5 and GPT-4o (few-shot)}: We add 20 examples (selected uniformly over engagement scores) from our training data in the prompt. 
\end{itemize}

Note that the RoBERTa and GPT baselines (text models) rely solely on the text embeddings of queries, while the other methods depend only on the non-text features we proposed in Section~\ref{sec:feature}. For RoBERTa, we added a linear layer to transform the CLS embedding into a continuous score in [0, 1].
For GPT baselines, we prompted it to score input keywords with scores in [0, 1], which are then used for ranking. The prompt is described in \Cref{appendix_prompts}.

\section{Results}
\label{sec:rs}

\begin{table*}[htbp]
\centering
\resizebox{0.7\textwidth}{!}{\begin{tabular}{lcccccc}
\toprule
\textbf{Method} & \textbf{Hit@5} & \textbf{Hit@50} & \textbf{Hit@100} & \textbf{Hit@500} & \textbf{Kendall's Tau} & \textbf{MSE} \\ \midrule
Frequency & 0.00 & 0.04 & 0.05 & 0.17 & 0.34 & - \\ \midrule
XGBoost (all features) & \textbf{0.20} & \textbf{0.42} & \textbf{0.50} & \textbf{0.69} &\textbf{0.52} & \textbf{0.0038} \\
XGBoost (behavioral only) & 0.00 & 0.26 & 0.39 & 0.63 & 0.50 & 0.0041 \\
XGBoost (financial only) & 0.00 & 0.18 & 0.17 & 0.55 & 0.43 & 0.0049 \\
XGBoost (catalog only) & 0.00 & 0.08 & 0.14 & 0.46 & 0.17 & 0.0063 \\  \midrule
Random Forest & \textbf{0.20} & 0.36 & 0.37 & 0.67 & 0.51 & 0.0040 \\
Lasso & 0.00 & 0.34 & 0.34 & 0.62 & 0.46 & 0.0043 \\
Ridge & 0.00 & 0.30 & 0.32 & 0.60 & 0.47 & 0.0044 \\
Elastic Net & 0.00 & 0.34 & 0.32 & 0.63 & 0.47 & 0.0043 \\
Linear & 0.00 & 0.32 & 0.31 & 0.61 & 0.47 & 0.0043 \\ \midrule
RoBERTa & \textbf{0.20} & 0.28 & 0.34 & 0.50 & 0.32 & 0.0112 \\
GPT-3.5 (zero-shot) & 0.00 & 0.06 & 0.11 & 0.30 & 0.05 & 0.4024 \\
GPT-3.5 (few-shot) & 0.00 & 0.06 & 0.08 & 0.30 & 0.05 & 0.1049 \\
GPT-4o (zero-shot) & 0.00 & 0.16 & 0.18 & 0.40 & 0.14 & 0.3822 \\
GPT-4o (few-shot) & 0.00 & 0.14 & 0.14 & 0.41 & 0.12 & 0.0486 \\
\bottomrule
\end{tabular}}\caption{The results of query ranking and engagement prediction of different methods.}\label{t:rs}
\end{table*}

\subsection{High Consideration Query Prediction}
\label{rs:engagement}

\Cref{t:rs} shows the query prediction results for all methods.

\textbf{Decision Tree Ensembles yield best overall performance.}
The gradient-boosted XGboost ensemble achieves the best performance across all metrics, with Random Forest achieving similar results. These ensemble-based models outperform all other single models, due to robustness and ability to capture diverse patterns within the data. 
Among linear methods, Lasso regression exhibits the best performance, although the difference is not statistically significant when compared to Elastic Net. We observe that the Hit$@k$ results of the frequency-based ranking baseline are notably poor. 
This verifies our claim in Section~\ref{sec:intro} that query frequency information does not necessarily correlate with customer consideration. 

\textbf{We achieve high recall for the top queries.}
When measuring recall of the top 500 HC queries with Hit$@500$, around 70\% of them were identified by XGBoost and approximately 60\% were identified by all other models trained with our proposed features. The robust performance across models highlights the reliability and consistency of our approach in accurately identifying HC queries.

\textbf{Ranking over the entire test set is accurate.}
When measuring ranking performance on the entire test set with Kendall's Tau, XGBoost and Random Forest both show a strong correlation with the ground truth rankings.
Rankings from the regression models show moderate correlations. We also observe that in general MSE follows a similar trend with Kendall's Tau. This is expected since both metrics consider the entire test set. However, when Kendall's Tau indicates low correlation, the MSE difference between two methods can be large (comparing zero-shot and few-shot GPT baselines).

\textbf{Text-based methods underperform feature-based models.} This is unsurprising given that those methods are typically pre-trained to capture semantic similarities in text.
Even GPT-3.5 performs poorly in all metrics, with little or no improvement even with few-shot in-context learning. GPT-4o achieves better results than GPT-3.5 but still cannot beat all other feature-based models. 
The task of EQR requires predicting on unseen queries, which may lead to challenges in generalization when facing queries that are dissimilar in semantics to those encountered during (pre-)training. 
This shows that query semantics alone are not sufficient for this task; behavioral features provide stronger cues.

\begin{figure*}[ht!]
    \centering
    \begin{subfigure}[b]{0.33\textwidth}
    \centering
    \includegraphics[width=\textwidth]{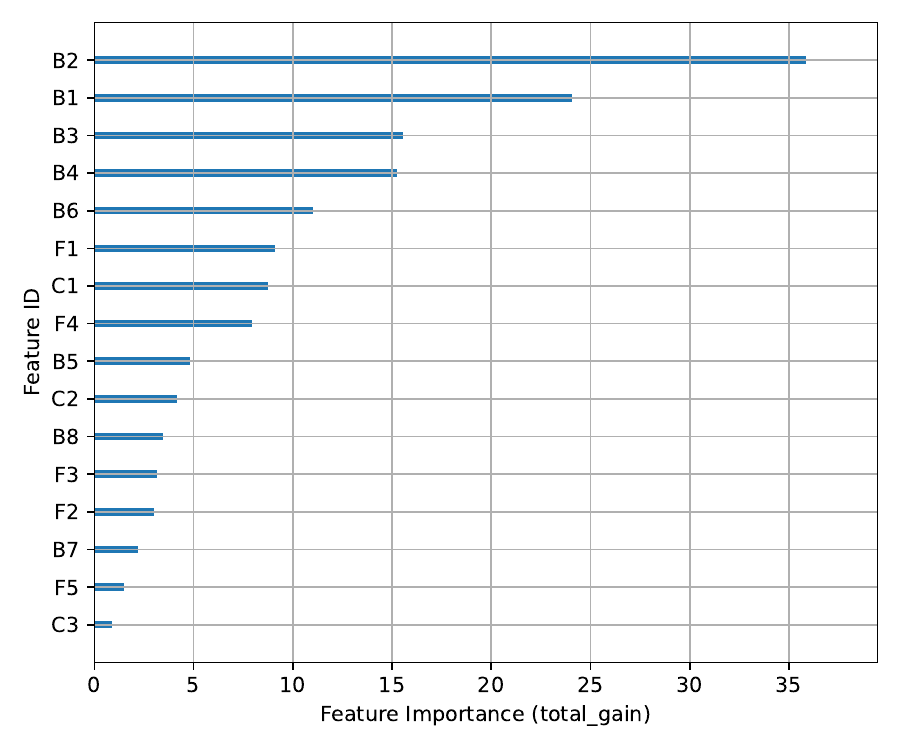}
    \caption{Feature importance by total gain.}
    \label{fig:importance-gain}
    \end{subfigure}%
    \begin{subfigure}[b]{0.33\textwidth}
        \centering
        \includegraphics[width=\textwidth]{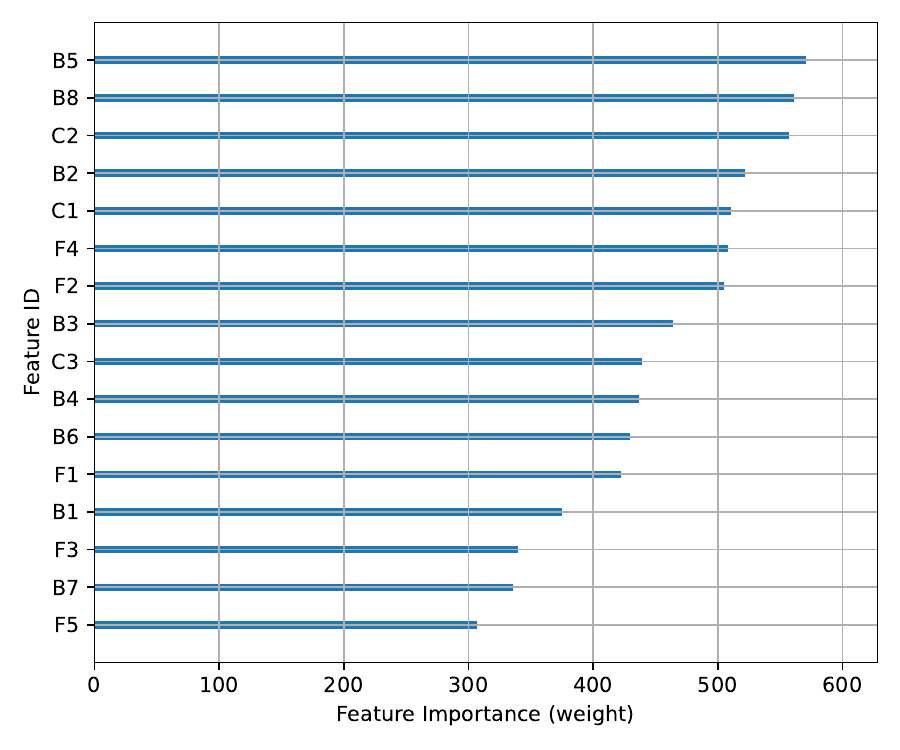}
        \caption{Feature importance by weight.}
        \label{fig:importance-weight}
    \end{subfigure}%
    \begin{subfigure}[b]{0.33\textwidth}
        \centering
        \includegraphics[width=\textwidth]{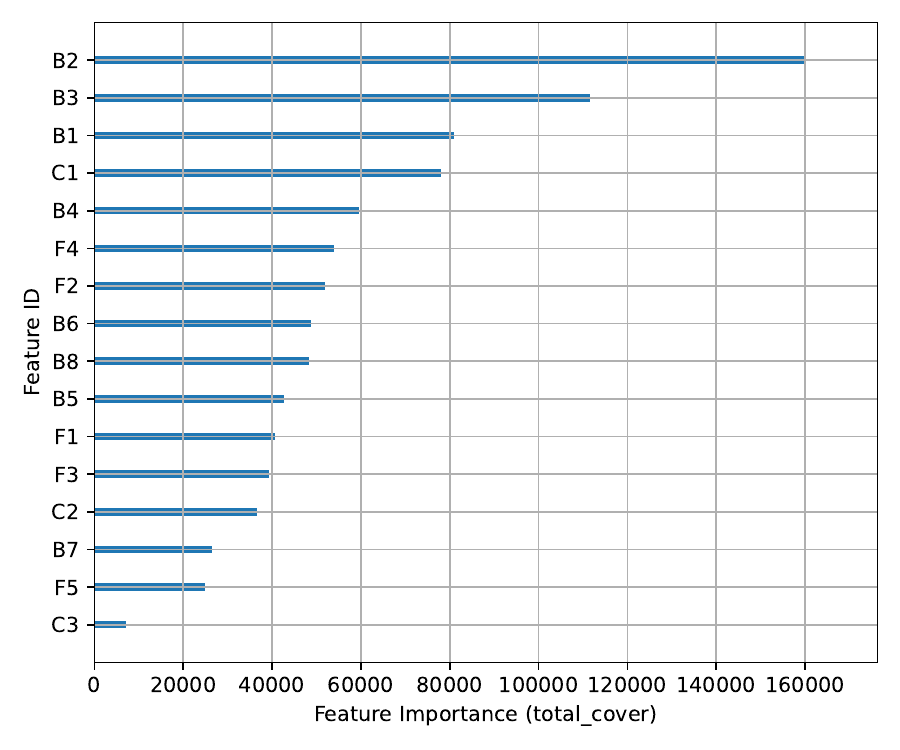}
        \caption{Feature importance by total cover.}
        \label{fig:importance-cover}
    \end{subfigure}
    \caption{Feature importance vales obtained from XGBoost.}
    \label{fig:importance}
\end{figure*}

\begin{table}[ht!]
\resizebox{0.85\columnwidth}{!}{
\begin{tabular}{p{2cm}p{5cm}}
\hline
\textit{High \newline Consideration} & \texttt{\mbox{iphone pro max}, \mbox{shark matrix},  \mbox{fujifilm x100f}, 
\mbox{hard drive ssd}, 
\mbox{bose tv speaker soundbar}, 
\mbox{tv 4k}, \mbox{xiaomi scooter}, \mbox{nikon d500},  \mbox{dji mini} , \mbox{vacuum cleaners}
} \\
\hline
\textit{Low \newline Consideration} & \texttt{\mbox{rubber mats for gym}, \mbox{pink belt}, 
\mbox{purse strap}, 
\mbox{keychain wristlet}, 
\mbox{foldable shoes}, \mbox{mushroom lamp},
\mbox{paw patrol toys},
\mbox{toys for boys}, 
\mbox{iridescent earrings}} \\
\hline
\end{tabular}}
\caption{Examples of top and bottom ranked queries.}
\label{t:query}
\end{table}

\subsection{Ablation Study and Feature Importance}

We conduct an ablation study on the feature groups from \Cref{sec:feature}.
Specifically, we examine how our best XGBoost model's performance changes when using each feature group independently. 
The results are shown in Table~\ref{t:rs} (row 3 to row 5). 
We observe a significant drop in performance across metrics when utilizing only one group of features. 
This drop indicates that behavioral features are of the most importance, followed by financial features. 

In \Cref{fig:importance} we plot the importance of different features obtained from XGBoost calculated with three different methods: (1)\textit{Total Gain} sums up the total gain achieved by using the feature across all splits (i.e., the reduction in entropy achieved by the split), reflecting its contribution to improving the model's performance; 
(2)\textit{Total Covererage} represents the total number of samples that the feature splits across all trees; 
and (3)\textit{Weight} measures the frequency of a feature's use in splitting the data across all trees. 
Though the interpretation of feature importance can be different, it is consistent across the three methods that the top two features are all behavior related, which aligns with the results in \Cref{t:rs}.

\subsection{Human Evaluation}
\label{sec:humaneval}

To validate the precision of our method in identifying HC queries, we used our model to make predictions on a large sample of unseen traffic in terms of our QA widget. 
We then created a set of 1,500 queries by selecting evenly from both the highest and lowest ranked queries in terms of predicted engagement scores. 
Expert annotators judged the shuffled set and classified each query as HC or not. 
The human labels were used as ground truth, and the model's top ranked queries were considered HC. 
We computed the precision of our model as 96\%, indicating its ability to generalize to unseen queries and potentially replace human annotators.

\section{Commercial Deployment}

Our model has been deployed in a production setting to identify HC queries for more than one year. 
As an initial step towards commercial deployment, we performed a head-to-head comparison between queries chosen by human experts and our model. Each group selected 500 HC queries, for which we curated and deployed high-quality QA content. As our metric, real engagement metrics (\cref{eq:y}) were measured over a 30-day period. Results showed that the model-chosen queries outperformed the human-selected set with a relative increase of 6\%.

Having validated the precision of our predictions (\S\ref{sec:humaneval}) and their downstream impact on customers, the model was moved to full commercial deployment. 
Removing the need for human annotators enables scaling HC query selection applications from an order of thousands to millions with relative ease. This, in turn, enables rapid model improvements by building an engagement-based feedback loop for the model to learn from its own predictions.

In Table~\ref{t:query}, we show some sampled queries from the top 10\% and bottom 10\%, as determined by the predicted rankings of a random sample of queries.
We observe that a significant portion of top queries are related to various types of electronics where customers greatly benefit from curated QA content or articles when making a purchase decision. Conversely, lower-ranking queries  like ``rubber mats for gym'' typically involve products where specific knowledge is not essential for decision-making. 

\section{Discussion and Conclusion}

We introduce the task of Engagement-based Query Ranking in order to select High Consideration queries. We proposed three categories of features to train pointwise rankers to address this task. Our experimental results show that our proposed method achieves better performance than the baselines. The human evaluation indicates that our method could serve as an effective tool to save resources spent on error-prone human annotations.

One limitation of our work is the difficulty in accurately measuring the true recall of our model. Future work could consider the combination of product category predictions for queries and conduct the selection of high consideration queries for each product category. Another limitation of our work is that we do not consider optimizing the curated content for selected queries and rely on human experts to decide what content to be create. Furthermore, we did not address the removal of nearly duplicated queries, which would require a separate processing pipeline. 
Future work could focus on personalized content selection for HC queries, and leverage active learning techniques to optimize the content-level engagement metrics. Another promising direction to explore is the integration of behavioral and financial features, along with an innovative LLM-based optimization approach~\cite{senel-etal-2024-generative} to improve overall performance. 

\clearpage

\bibliography{acmart}

\clearpage
\onecolumn
\appendix

\section*{Appendix}
\section{Prompt Details}
\label{appendix_prompts}

The only difference between zero-shot and few-shot prompts are the presence of 20 additional pairs of (query, $e_{q}$) few-shot example. However due to legal and privacy reasons, we only include our zero-shot prompt below for generating engagement scores with reasons.
\\

\begin{figure*}[!hb]
    \centering
    \includegraphics[width=0.95\textwidth]{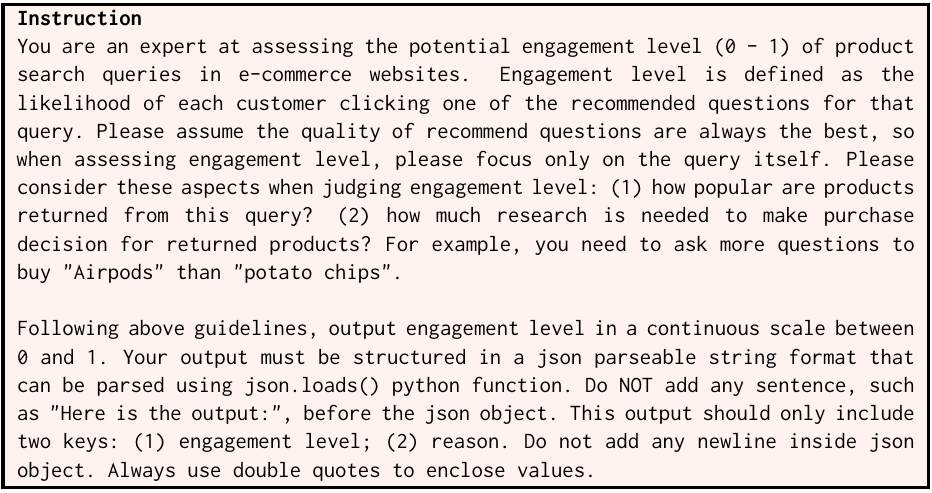}
    \caption{Zero-shot prompt used for generating engagement score and reasons using OpenAI GPT models.}
    \label{tab:openai_few_shot_prompt}
\end{figure*}

\end{document}